\documentclass[prd,aps,twocolumn,a4paper,nofootinbib,showpacs]{revtex4-1}

\newif\ifusesec
\usesectrue % or
\usepackage{graphicx,psfrag}
\usepackage{mathrsfs}
\usepackage{amsmath,amsfonts,amssymb}
\usepackage{multirow}
\usepackage{comment}
\usepackage{bm}

\DeclareSymbolFontAlphabet{\mathrsfs}{rsfs}
\DeclareMathAlphabet\mathbfcal{OMS}{cmsy}{b}{n}

\newcommand{\be}{\begin{equation}}
\newcommand{\ee}{\end{equation}}
\newcommand{\bea}{\begin{eqnarray}}
\newcommand{\eea}{\end{eqnarray}}
\newcommand{\bel}{\begin{align}}
\newcommand{\eel}{\end{align}}

\newcommand{\scri}{{\mathrsfs{I}}}

\def\i{{\rm i}}

\def\GMc2{G M_{\odot} c^{-2}}

\def\M{{\mathsf M}}

\def\mx{{\rm max}}

\def\F{{\cal F}}
\def\hF{\hat{\cal F}}
\def\lm{{\ell m}}

\def\lm{{\ell m}}

\def\lm{{\ell m}}

\def\m{{\mathsf m}}

\def\F{{\cal F}}

\DeclareSymbolFontAlphabet{\mathrsfs}{rsfs}
\DeclareMathAlphabet{\mathcal}{OMS}{cmsy}{m}{n}

\usepackage{color}

\definecolor{cyan}{rgb}{0,0.9,0.9}
\definecolor{orange}{rgb}{0.9,0.5,0}
\definecolor{magenta}{rgb}{1,0,1}
\definecolor{purple}{rgb}{0.8,0.4,0.8}
\definecolor{gray}{rgb}{0.8242,0.8242,0.8242}

\begin{document}

\title{Gravitational recoil in nonspinning black-hole binaries: The span of test-mass results}

\author{Alessandro \surname{Nagar}}

\affiliation{Institut des Hautes Etudes Scientifiques, 91440 Bures-sur-Yvette, France}
\affiliation{Department of Physics, University of Torino, 10125 Torino, Italy}                                                                                                                                                                   
\date{\today}

\begin{abstract}
We consider binary systems of coalescing, nonspinning, black holes of masses $m_{1}$ and $m_{2}$ and
show that the gravitational recoil velocity for any mass ratio can be obtained accurately  by  extrapolating
the waveform of the test-mass limit case. The waveform obtained in the limit  $m_1/m_2\ll 1$ via a perturbative
approach is extrapolated in $\nu= m_{1} m_{2}/(m_{1}+m_{2})^{2} $ multipole by multipole using 
the corresponding, analytically known, leading-in-$\nu$ behavior. The final kick velocity computed from this $\nu$-flexed
waveform is written as $v(\nu)/c = 0.04457 \nu^2\sqrt{1-4\nu}\,(1-2.07106\nu + 3.93472\nu^2 -4.78404\nu^3+2.52040\nu^{4})$
%(1 -2.0641 \nu + 3.7666\nu^2-3.6050\nu^3)$
and is compatible with the outcome of numerical relativity simulations
\end{abstract}


\date{\today}

\maketitle

\section{ Introduction}
Interference between the multipoles of the gravitational waves (GW) emitted from coalescing black-hole binaries 
of masses $m_1$ and $m_2$ carries away linear momentum and thus imparts a recoil to the final merged black hole. 
The accurate calculation of this recoil velocity, also referred as kick, has been the topic of analytical and 
numerical studies in recent years~\cite{Damour:2006tr,Sopuerta:2006wj,Schnittman:2007ij,Baker:2006vn,Gonzalez:2006md,
Gonzalez:2008bi,Campanelli:2007cga,LeTiec:2009yg,Lousto:2011kp,Lousto:2012gt,Buchman:2012dw}.
In particular, after assessing the properties of the kick velocity for nonspinning black-hole binaries, 
numerical relativity (NR) went on to investigate the effect the black-hole spins have on the final kick. 
The most interesting and astrophysically relevant result is that high recoil velocities, 
of about a few thousands of km/s, can be reached for nonaligned spin 
configurations~\cite{Lousto:2011kp,Lousto:2012gt}.

When one black hole is much more massive than the other, 
$\M\equiv m_2\gg \m\equiv m_1$ ($\m/\M\equiv 1/q\ll1$), the kick is obtained 
from the GW emission computed using black hole perturbation theory~\cite{Bernuzzi:2010ty,Sundararajan:2010sr}.
When the larger black hole is nonspinning, Ref.~\cite{Bernuzzi:2010ty} used Regge-Wheeler-Zerilli (RWZ)
perturbation theory~\cite{Nagar:2005ea} to calculate the GW emission from the transition from inspiral 
to plunge of a point-particle source subject to leading-order (LO) analytical (effective-one-body), 
resummed radiation reaction force. When the larger black hole is spinning,~\cite{Sundararajan:2010sr} 
solved the Teukolsky equation with a point-particle source term subject to a numerical, adiabatic, 
radiation reaction force. In the nonspinning case, both studies essentially agreed on the value 
of the final recoil velocity: Ref.~\cite{Sundararajan:2010sr}
got $v/[c(\m/\M)^2]=0.044$, using up to $\ell=6$ multipoles, while Ref.~\cite{Bernuzzi:2010ty} estimated
$v/[c(\m/\M)^2]=0.0446$ using multipoles up to $\ell=8$. Reference~\cite{Sundararajan:2010sr}   
studied whether the perturbative result can be accurately extrapolated to any mass ratio 
using the $\nu$-scaling corresponding to the LO multipolar contribution~\cite{Fitchett:1984qn}
\be
\label{eq:Fitchett}
v(\nu)/c = 0.044 \nu^2 \sqrt{1-4\nu},
\ee
where $\nu=m_1 m_2/M^2$, with $M=m_1+m_2$, is the symmetric mass ratio.
It was found that this scaling is rather inaccurate when $\nu\sim 0.2$, 
as it predicts values that are larger by $\sim 50\%$ than the NR results.

In this paper  we show that extrapolating in $\nu$ the test-mass waveform multipole by multipole up
to multipole order $\ell=8$ and then computing the recoil from this $\nu$-flexed waveform, allows 
one to get an improved version of the LO scaling that is compatible with the NR 
results of Refs.~\cite{Gonzalez:2006md,Gonzalez:2008bi,Buchman:2012dw}.

\section{Extrapolating in ${\bm \nu}$ test-mass results}

Let us start by pointing out a systematic flaw in assuming the LO 
scaling~\eqref{eq:Fitchett}. The RWZ-normalized multipolar decomposition of
the waveform is (for equatorial motion)
\vspace{-3mm} 
\be
\vspace{-2.5mm}
h_+ - \i h_\times = \dfrac{1}{r}\sum_{\ell=2}^{\ell_{\max}}\sum_{m=-\ell}^{\ell}\sqrt{\dfrac{(\ell +2)!}{(\ell-2)!}}\;\i^{{\epsilon}} \Psi^{(\epsilon)}_{\lm}{}_{-2}Y^{\lm}(\theta,\phi),
\nonumber
\ee
where $\epsilon=0,1$ is the parity of $\ell+m$. The functions $\Psi^{(\epsilon)}_{\lm}\equiv\Psi_{\lm}^{(\epsilon)}(t;\nu)$,
(e.g., computed from a NR simulation), are normalized as in Ref.~\cite{Bernuzzi:2010ty}. 
In the perturbative context $(\nu\to 0)$,  they are a solution of the Zerilli ($\epsilon=0$) 
and Regge-Wheeler ($\epsilon=1$) equations with a point-particle source term~\cite{Nagar:2006xv,Bernuzzi:2010ty}.
The GW linear momentum flux in the equatorial plane is
\begin{align}
\label{eq:P_flux}
{\cal F}_x^{\bf P} + \i {\cal F}_y^{\bf P}= \dfrac{1}{8\pi}\sum_{\ell=2}^{\ell_{\max}}\sum_{m=-\ell}^\ell\i \bigg[a_{\lm}\dot{\Psi}_\lm^{(0)}\dot{\Psi}^{(1)*}_{\ell,m+1}\nonumber\\
+b_\lm\sum_{\epsilon=0,1} \dot{\Psi}^{(\epsilon)}_\lm\dot{\Psi}^{(\epsilon)*}_{\ell+1,m+1}\bigg],
\end{align}
where the numerical coefficients $(a_\lm,b_\lm)>0$ are given in Eqs.~(16)-(17) of~\cite{Bernuzzi:2010ty},
and $\Psi^*_{\lm} = (-1)^m\Psi_{\ell,-m}$. The  (complex) recoil velocity at time $t$ is obtained as
\be
\label{eq:v_kick}
v_x + \i v_y=-\dfrac{1}{M}\int_{-\infty}^{t}\left({\cal F}_x^{\bf P}+\i \F_y^{\bf P} \right) dt'.
\ee
For each multipole, the leading-in-$\nu$ (completely explicit) dependence is~\cite{Damour:2008gu}
$\Psi_\lm^{(\epsilon)}\propto \nu c_{\ell + \epsilon}(\nu) $,
where $c_{\ell +\epsilon}(\nu)\equiv X_2^{\ell + \epsilon-1} + (-)^m X_1^{\ell+\epsilon-1}$,
with $X_i = m_i/M$ so that $X_1+X_2=1$ and $X_1 X_2=\nu$. The convention 
we adopt here is $X_2 > X_1$,  i.e., $X_2-X_1=\sqrt{1-4\nu}$, so that $c_{\ell+\epsilon}(0)=1$.
The explicit $\nu$-dependence in Eq.~\eqref{eq:P_flux} comes as sum of products of $c_{\ell+\epsilon}(\nu)$.
Defining individual rescaled fluxes as  $\hF_{\ell m \ell'm'}\equiv  \i/(8\pi) \alpha_{\lm} \dot{\Psi}_{\lm}^{(\epsilon)}\dot{\Psi}_{\ell' m'}^{(\epsilon')*}/[\nu^2 c_{\ell+\epsilon}(\nu)c_{\ell'+\epsilon'}(\nu)]$ (with either $\alpha_{\lm}=a_{\lm}$ or $\alpha_{\lm}=b_{\lm}$), Eq.~\eqref{eq:P_flux} reads
\begin{align}
\label{eq:flux_partial}
&\F_x^{\bf P}+\i\F_y^{\bf P}= \nu^2 \sqrt{1-4\nu}\bigg\{\hF_{223-3}+\hF_{2-231}+\hF_{2-221}+\dots\nonumber\\
                              &+(1-3\nu)\hF_{334-4}+\dots +(1-3\nu)(1-2\nu)\hF_{445-5}+\dots \bigg\},
\end{align}
where we wrote just a few terms to indicate that the explicit (leading) $\nu$-dependence of the flux is more 
complicated than just the LO one.
%================================
% Recoil table: explicit comparison with NR
%================================
\begin{table}[t]
  \caption{\label{tab:CCC} Final recoil velocity: comparing the (multipolar) $\nu$-extrapolated 
    RWZ result, $v^{{\rm RWZ}_{\nu}}_{{\rm end}}$, the leading-order extrapolation, Eq.~\eqref{eq:Fitchett}, 
    $v^{{\rm RWZ}_{\rm LO}}_{\rm end}$ and the NR values of~\cite{Buchman:2012dw}. 
    As a conservative error estimate, the $v^{{\rm RWZ}_{\nu}}_{{\rm end}}$ can be larger by 1 to 2$\%$. See text for details.}
    \begin{center}
    \begin{ruledtabular}
      \begin{tabular}{ccccc}
        $q$   &   $\nu$  &    $v^{\rm NR}_{\rm end}$[km/s]  &   $v^{{\rm RWZ}_\nu}_{\rm end}$[km/s]&   $v^{{\rm RWZ}_{\rm LO}}_{\rm end}$[km/s] \\
        \hline 
        2      &  $0.\bar{2}$ & $148\pm 2$  &   151.3  & 219.9 \\
        3      &  $0.1875$    & $ 174\pm 6$ & 169.5   & 234.8 \\
        4      &  $0.1600$    &  $ 157\pm 2$ & 154.2  & 205.2 \\
        6      &   $ 0.1224$  &  $ 118\pm 6$  & 114.1  & 143.1        
         \end{tabular}
  \end{ruledtabular}
\end{center}
\end{table}
%----------------------------------------------------------------------------------
Let us consider now the $\nu\to 0$ gravitational waveform $\Psi_\lm^{(\epsilon)}(t;0)$ 
obtained solving the RWZ equations with a point-particle source subject to leading-order, 
resummed, analytical radiation reaction force. The mass ratio is $\m/\M=10^{-3}$. This waveform 
was computed in Ref.~\cite{Bernuzzi:2011aj} using the hyperboloidal layer approach~\cite{Zenginoglu:2009ey}, 
which allowed us to: i) extract waves at $\scri^+$; ii) obtain high-resolution data (the numerical error is not an issue).
The quasicircular inspiral starts at $r_0=7\M$. 
The recoil velocity obtained from Eq.~\eqref{eq:P_flux} with 
$\ell_{\max}=7$ is $v(0)/[c (\m/\M)^2] = 0.04457$, consistent 
with~\cite{Bernuzzi:2010ty}. Analyzing the  corresponding 
$\hat{\F}_{\ell m \ell' m'}(t;\,0)$'s ($\nu\equiv \m/\M$, $c_{\ell+\epsilon}(0)=1$), 
one finds that the (complex) coefficients of the different $\nu$-dependent
terms in the curly bracket of
Eq.~\eqref{eq:flux_partial} are essentially in phase. It follows that
the $\nu$-extrapolation of $v(0)$ done using Eq.~\eqref{eq:Fitchett} 
[i.e., ignoring the extra factors $(1-3\nu)$, $(1-3\nu)(1-2\nu)$ etc. in Eq.~\eqref{eq:flux_partial}]
is inaccurate (and in particular gives a value {\it larger} than the correct  one) at least
because the $\nu$ dependence of several subleading terms crucially contributing 
to the momentum flux is not taken into account correctly.
For example, for $\nu=0.2$, where the function $\nu^2 \sqrt{1-4\nu}$ 
gets its maximum, the values of the extra $\nu$-factors are $(1-3\times 0.2)=0.4$ and  
$(1-3\times 0.2)(1-2\times 0.2)=0.24$. The LO $\nu$-scaling is then incorrectly 
amplifying $\hat{\F}_{334-4}$ and $\hat{\F}_{445-5}$  by 2.5 and  $4$ times respectively.
%================
% Fig.
%================
\begin{figure}[t]
 \includegraphics[width=0.4\textwidth]{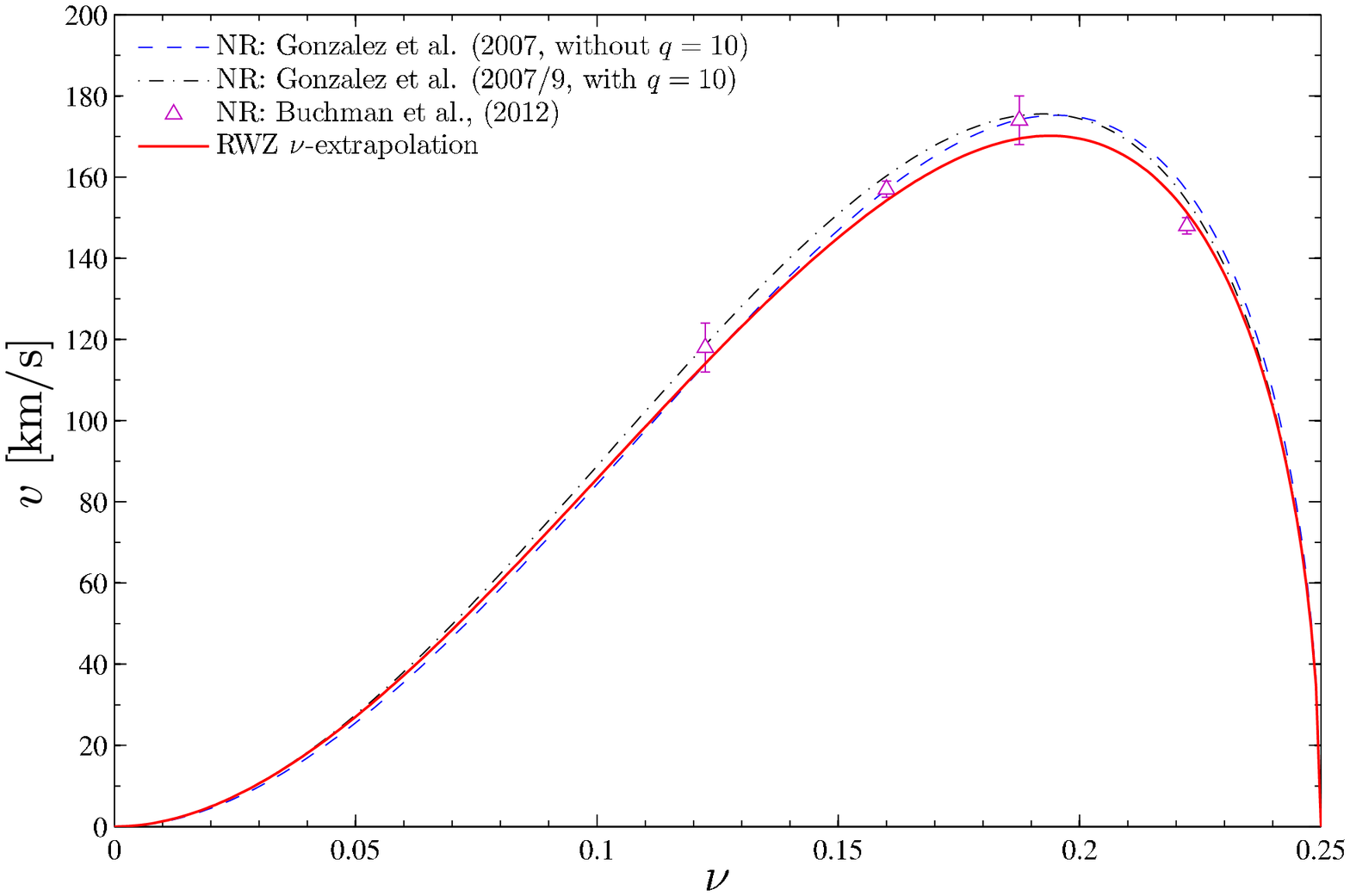}\\
 \includegraphics[width=0.4\textwidth]{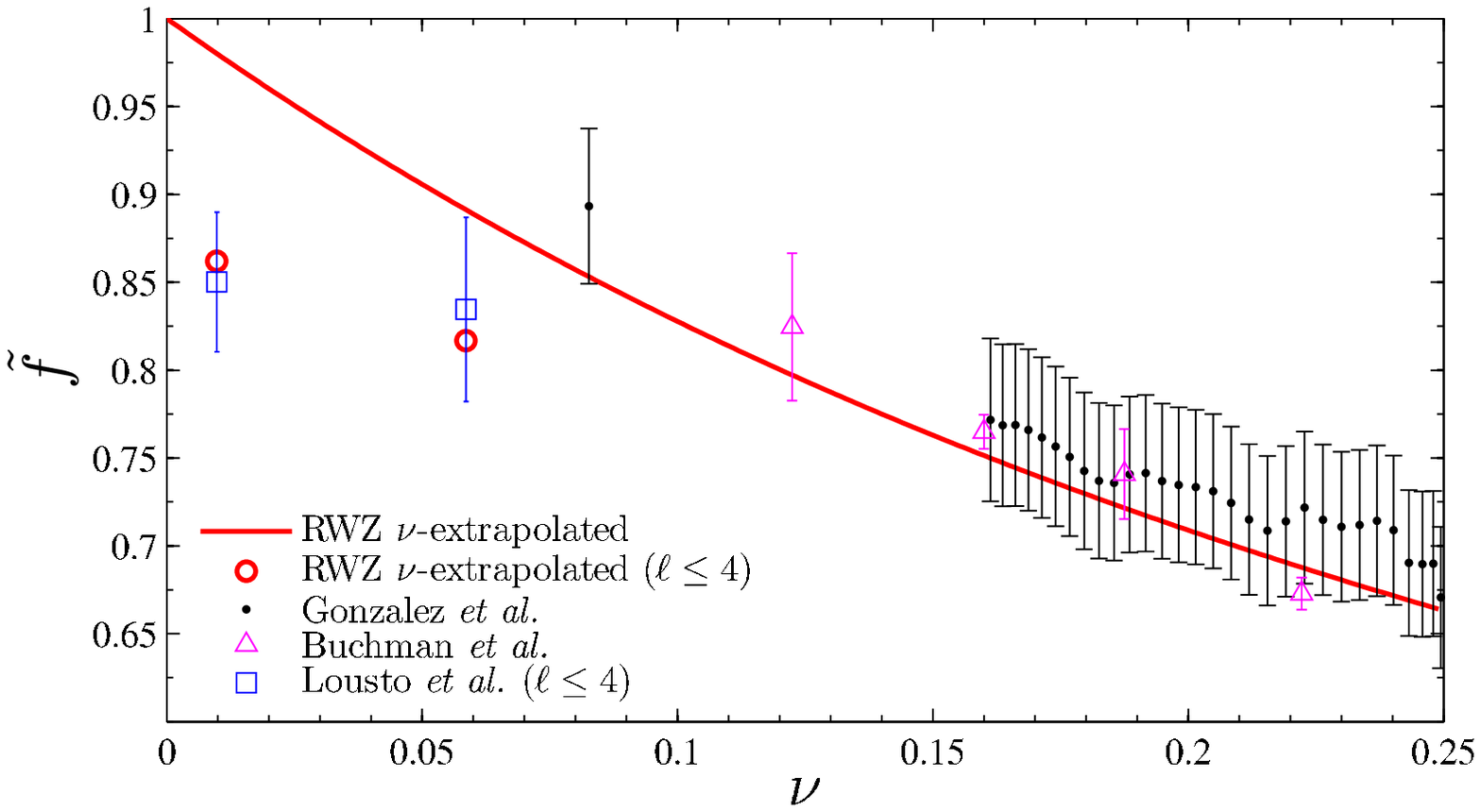}\\
    \caption{ \label{fig:fig_kick} (color online) Top: magnitude of the final recoil velocity versus $\nu$. 
     The data points of~\cite{Buchman:2012dw} and the fits to the NR data of Refs.~\cite{Gonzalez:2006md,Gonzalez:2008bi}  
     are compared with the result of the extrapolation in $\nu$ of the RWZ multipolar waveform (red curve). 
     Bottom: the extrapolated reduced function $\tilde{f}(\nu)\equiv v(\nu)/[v(0)\nu^2\sqrt{1-4\nu}]$
     contrasted with the actual NR data of~\cite{Gonzalez:2006md,Gonzalez:2008bi,Lousto:2010qx,Lousto:2010ut,Buchman:2012dw}.} 
\end{figure}
%=========================================================================

To extrapolate in $\nu$ the multipolar waveform, we take $\hat{\Psi}^{(\epsilon)}_{\lm}(t;\,0)\equiv \Psi^{(\epsilon)}_{\lm}(t;\,0)/(\m/\M)$, 
multiply it by the corresponding leading-order $\nu$ dependence, so to get the $\nu$-dependent
function (addressed as  RWZ$_\nu$ in the following) $\Psi_{\lm}^{(\epsilon)}(t;\,0_\nu)\equiv \nu c_{\ell+\epsilon}(\nu)\hat{\Psi}^{(\epsilon)}_{\lm}(t;\,0)$. [The notation $0_\nu$ is a reminder that only the leading order $\nu$ dependence 
of each multipole is included and so $\Psi^{(\epsilon)}_\lm(t;\,0_\nu)\neq \Psi^{(\epsilon)}_\lm(t;\,\nu)$].
Using $\Psi_\lm^{(\epsilon)}(t;\,0_\nu)$ in Eq.~\eqref{eq:P_flux} we get the linear
momentum flux versus time and then the kick velocity via Eq.~\eqref{eq:v_kick}. 
Since the waveform starts at time $t_0>-\infty$, the boundary 
condition  $M v_0\equiv -\int_{-\infty}^{t_0}(\F^{\bf P}_x+\i\F^{\bf P}_y)dt$ in Eq.~\eqref{eq:v_kick}
is fixed as the center of the velocity hodograph during the inspiral~\cite{Bernuzzi:2010ty}.

Table~\ref{tab:CCC} compares the final kick velocity $v\equiv |v_x+\i v_y|$ obtained from
 the RWZ$_\nu$ waveform with the most recent NR calculations~\cite{Buchman:2012dw}, 
using the SpEC~\cite{Scheel:2008rj} code, with $q=(2,3,4,6)$ (and retaining only multipoles with $\ell\leq6$). 
The extrapolated values are very close to the NR ones, in two cases within their error bars. 
By contrast, the last column of the table highlights how inaccurate the leading-order scaling is. 
The uncertainty on the RWZ$_\nu$ values has essentially two sources: (i) the fact that $\m/\M\ll 1$, 
but always $\m/\M\neq 0$ and (ii) the effect of multipoles selected by the condition $\ell_{\rm max}> 7$.
In Table~III of Ref.~\cite{Bernuzzi:2010ty} it was shown that changing $ \m/\M=10^{-3}$ to $\m/\M=10^{-4}$ 
was increasing the final kick by $\sim 0.5\%$. In addition, we checked that the relative difference 
between taking  $\ell_{\rm max}=6$  $[v(0) (\M/\m)^2 = 0.04383 ]$ and $\ell_{\rm max}=7$ ($v(0)(\M/\m)^{2}=0.04457$) 
is  as large as $\sim 1.7\%$ when $\m/\M=10^{-3}$, but becomes as small as $10^{-3}$ 
for $q=6$ and $10^{-4}$ for $q=2$.  As a conservative error estimate, the extrapolated values 
of Table~\ref{tab:CCC} can be {\it larger} by $1$ to $2\%$.

Figure~\ref{fig:fig_kick} compares $v(\nu)$ with $0\leq \nu \leq 0.25$ (solid curve, red online) 
with available fits obtained from the comprehensive numerical study of 
Refs.~\cite{Gonzalez:2006md,Gonzalez:2008bi}. We also show the data of Ref.~\cite{Buchman:2012dw}. 
The data of Refs.~\cite{Gonzalez:2006md,Gonzalez:2008bi} are represented by two different fits: 
$v^{\rm NR}=1.20\times 10^{4}\nu\sqrt{1-4\nu}(1-0.93\nu)$ (dashed, blue online), proposed in Ref.~\cite{Gonzalez:2006md} 
without including the $q=10$ data of~\cite{Gonzalez:2008bi}, and $v^{\rm NR }/c= 0.04396\nu^2\sqrt{1-4\nu}(1-1.3012\nu)$,
with $c=299792.458$~km/s (dot-dashed) done in~\cite{Bernuzzi:2010ty} 
including the $q=10$ data. The maximum value of the RWZ$_{\nu}$ curve is $v_{\rm max}=170.164$~km/s
(at $\nu=0.194$), quite close to $v_{\max}^{\rm NR}=175.2\pm 11$~km/s computed in~\cite{Gonzalez:2006md}.
A more precise quantitative information is given by (bottom panel of Fig.~\ref{fig:fig_kick}) 
the normalized quantity $\tilde{f}=v(\nu)/[v(0)\nu^2\sqrt{1-4\nu}]$ obtained 
from the  extrapolated $v(\nu)$ (solid line). For completeness, we also exhibit the raw
NR data of Refs.~\cite{Gonzalez:2006md,Gonzalez:2008bi,Buchman:2012dw} as well
as those of Refs.~\cite{Lousto:2010qx,Lousto:2010ut} for the challenging values 
$q=15$ and $q=100$,  the highest simulated so far. Note that for these $q$'s the 
recoil velocity is systematically underestimated since the multipoles with  $\ell > 4 $ 
were neglected in Refs.~\cite{Lousto:2010qx,Lousto:2010ut}. Notably,  if the 
extrapolation is done retaining {\it only} the multipoles with $\ell \leq 4$, the RWZ$_{\nu}$ 
result for $q=15$ and $q=100$ (red circles in the bottom panel of Fig.~\ref{fig:fig_kick}) 
is compatible with the NR points.
The complete RWZ$_{\nu}$ $\tilde{f}(\nu)$ curve is accurately fitted ($\Delta\tilde{f}\equiv \tilde{f}-\tilde{f}^{\rm RWZ_{\nu}}\sim 10^{-5}$) 
by the quartic trend $\tilde{f}(\nu)= 1-2.07106\nu + 3.93472\nu^2 -4.78404\nu^3+2.52040\nu^{4}$.
[A cubic trend yields instead $\tilde{f}(\nu)= 1-2.06407\nu + 3.76663\nu^2 -3.60498\nu^3$ with $\Delta\tilde{f}\sim 10^{-4}$, 
undistinguishable on the scale of Fig.~\ref{fig:fig_kick}. Note that the (less accurate) quadratic trend was instead 
suggested in both Ref.~\cite{Damour:2006tr} using the effective-one-body formalism and Ref.~\cite{Sopuerta:2006wj} using the 
close-limit approximation].  It would be interesting to extract  $\tilde{f}(\nu)$ accurately from ad hoc NR simulations.
%==========
% Fig: antikick
%==========
\begin{figure}[t]
 \includegraphics[width=0.4\textwidth]{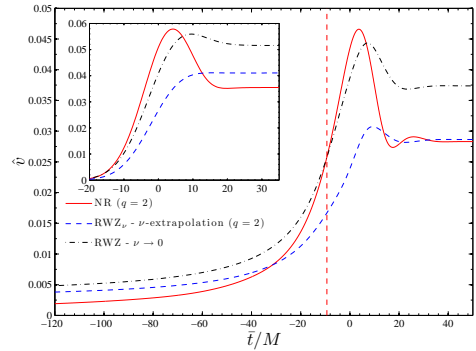} \\
  \caption{ \label{fig:antikick} (color online) Time evolution of $\hat{v}\equiv v/(c \nu^2\sqrt{1-4\nu})$ for $q=2$  
  obtained from (a restricted sample of) multipoles of the NR waveform and from the $\nu$-extrapolated 
  RWZ$_\nu$ ones.   The vertical line indicates the NR merger.  Inset: corresponding analytical approximations, 
  Eq.~\eqref{eq:vt}, to $\hat{v}(t)$. The nonextrapolated $\nu\to 0$ curves are also shown for completeness.}
\end{figure}
%----------------------------------------

%====================
% Magnitude of the antikick
%====================
\begin{table}[t]
  \caption{\label{tab:CCC_antikick} Final and maximal recoil velocity computed from the NR (boldface) 
  and RWZ$_\nu$ $\nu$-extrapolated waveform for a restricted sample of waveform multipoles ($\ell,m)$ with 
  $m=\ell$ up to $\ell=6$, (2,1) and (3,2). Here it is $\hat{v}\equiv v/(c\nu^2\sqrt{1-4\nu})$.}
  \begin{center}
    \begin{ruledtabular}
	\begin{tabular}{cccccc}
	   $q$ & $v_{\rm end} $[km/s]& $\max(v)$[km/s] & $\hat{v}_{\rm end} $ & $\max(\hat{v})$ & $\Delta\hat{v}$   \\
	   \hline
	   2 & {\bf 139.60} & {\bf 229.94}      & {\bf 0.0283}  &  {\bf 0.0466} & {\bf 0.0183}\\
	      &        141.32  &        151.72        &       0.0286   &  0.0307 & 0.0029\\
	   3 &  {\bf 162.04}& {\bf 243.74}       & {\bf 0.0308} &  {\bf 0.0462}  & {\bf 0.0154}\\
	      &         156.70 &        170.58 \      &        0.0297  &   0.0324 & 0.0026\\
	   4 &   {\bf 147.80}& {\bf 210.04}     & {\bf 0.0321} & {\bf 0.0456} & {\bf 0.0135}\\
 	      &         141.20 &155.49               & 0.0307  & 0.0338 & 0.0031\\
	   6 &   {\bf 107.80}& {\bf 144.17}   & {\bf 0.0336}  & {\bf 0.0449} & {\bf 0.0113}\\ 
	       &          102.82 & 115.12           & 0.0320  & 0.0358 & 0.0038 \\
	   \hline
	  $\infty$  & \dots & \dots  & 0.0374 &  0.0443 & 0.0070
         \end{tabular}
  \end{ruledtabular}
\end{center}
\end{table}
{\it Time evolution of kick velocity. --} We investigate now if the $\nu$-extrapolation is able to reproduce 
the structure of the well-known (post-merger) local maximum of $v(t)$, predicted and analytically 
explained in~\cite{Damour:2006tr} (see also~\cite{Price:2011fm}) and now known 
as ``antikick''~\cite{Schnittman:2007ij,Rezzolla:2010df}.
Since this information is not given in~\cite{Buchman:2012dw}, we have to compute $v^{\rm NR}(t)$  
from the (limited) number of NR $(\ell,m)$ waveform multipoles of~\cite{Buchman:2012dw} to which we have access. 
For both NR and RWZ$_\nu$ we use $\Psi_{\lm}^{(\epsilon)}$ with $m=\ell$ up to $\ell=6$ plus (2,1) and (3,2).
Table~\ref{tab:CCC_antikick} lists the final and maximum velocity obtained  from NR (boldface) 
and RWZ$_\nu$ data (cf. with Table~\ref{tab:CCC}), together with the magnitude of the 
antikick, $\Delta\hat{v}\equiv \max(\hat{v})-\hat{v}_{\rm end}$, with $\hat{v}\equiv v(t)/(c\nu^2\sqrt{1-4\nu})$.
Even with a limited number of multipoles, the $\nu$-extrapolated $v_{\rm end}$ is accurate; 
by contrast, the extrapolated antikick is much smaller than the corresponding NR one. 
The table is complemented  by the main panel of Fig.~\ref{fig:antikick}, where we contrast the $q=2$ $\hat{v}(t)$
for both NR and RWZ$_\nu$ data  (the original $\nu\to0$ curve is also added for completeness). 
Note that $\hat{v}(t)$ is plotted versus $\bar{t}\equiv t-t_{\rm max}$, where $t_{\rm max}$ corresponds 
to the maximum of  $\F_{\bf P}\equiv |\F_x^{\bf P}+\i \F_y^{\bf P}|$.  The vertical line indicates the
NR merger, defined as the peak of $|\Psi_{22}|$.

%==================================
% Numerical information around peak
%==================================
\begin{table}[t]
  \caption{\label{tab:CCC_max}Characterization of  $\max(\F_{\bf P})$ for the NR (boldface)
  and RWZ$_\nu$ waveforms (with a restricted sample of dominant multipoles). 
  Here is $\tilde{\bf \F}_{\bf P}^{\rm max}\equiv \F_{\bf p}^\mx/\nu^2\times 10^3$. The analytical estimate $v^{\rm end}_A$ 
  of the final recoil velocity (last two columns) is obtained from Eq.~\eqref{eq:vend}.} 
  \begin{center}
    \begin{ruledtabular}
             \begin{tabular}{ccccccc}
        $q$   & $\tilde{\F}_{\bf p}^\mx$   &  $\tau_\mx$ &  $Q$ &   $\epsilon_\mx $ & $v^{\rm end}_{\rm A}[{\rm km/s}]$ & $\hat{v}^{\rm end}_{\rm A}$  \\
        \hline 
        2      & ${\bf 3.009}$  & {\bf 7.505} & {\bf 1.770} & {\bf 0.011} & {\bf 174.85} & {\bf 0.0354}  \\ 
                 & $     1.463$    & 7.780  & 1.298 & -0.486 & 202.57 & 0.0410 \\
        %\vspace{1mm} \\
        3     & ${\bf 4.22}$   & {\bf 7.485} & {\bf 1.666} & {\bf -0.028} & {\bf 208.47} & {\bf 0.0396} \\
               & $2.330$   & 7.823 & 1.319 & -0.465 & 224.30 & 0.0426 \\
        %\vspace{1mm} \\
        4      & ${\bf 4.816}$   & {\bf 7.526} & {\bf 1.607} & {\bf -0.065} & {\bf 192.39} & {\bf 0.0418}\\ 
                & $2.930$   & 7.858 & 1.335 & -0.447 & 201.621 & 0.0438 \\ 
        %\vspace{1mm} \\
        6      & ${\bf 5.347}$   & {\bf 7.689} & {\bf 1.552} & {\bf -0.136} & {\bf 141.29} & {\bf 0.0440}\\ 
                & $3.730$ &   7.905 & 1.356 & -0.422 & 146.07 & 0.0455\\ 
               \hline
         $\infty$  & $6.499$  & 8.043 & 1.418 & -0.330 & \dots & 0.0516 
         \end{tabular}
  \end{ruledtabular}
\end{center}
\end{table}
%---------------------------------------------------------------------------------------------------------------------------------------------
\section{Discussion}
The results presented so far are consistent with the analytical explanation of the structure of 
the gravitational recoil given in Ref.~\cite{Damour:2006tr}. Essentially, Ref.~\cite{Damour:2006tr}
argued that the properties of $v(t)$ after the maximum of $\F_{\bf P}$ are approximately 
determined by what happens close to the peak of ${\bf \F}_{\bf P}$.
At time $t$ we have the complex integral~\eqref{eq:v_kick},
i.e. $v_x +\i v_y= \i {\cal I} = \i \int_{-\infty}^{t} \F_{\bf P}(t) e^{i\varphi(t)}dt$. 
Due to the {\it nonadiabatic} character of the evolution of the momentum flux, 
this integral is dominated by what happens near $\max[\F_{\bf P}(t)]$. 
Expanding around $t_{\max}$ one gets~\cite{Damour:2006tr}
\be
\label{eq:vt}
v_x+\i v_y\simeq \i \F_{\bf P}^{\rm max}e^{\i \varphi_{\max}}\sqrt{\dfrac{\pi}{2\alpha}}e^{\beta^2/(2\alpha)}{\rm erfc}(z),
\ee
with $z=-\sqrt{\alpha/2}(\bar{t}-\beta/\alpha)$, where $\alpha\equiv 1/\tau_{\rm max}^2(1-\i\epsilon_{\rm max})$
and $\beta=\i Q/\tau_{\rm max}$. Here $\tau^2_{\rm max}\equiv -\F_{\bf P}^{\rm max}/(d^2\F_{\bf P}/d\tau^2)^{\rm max}$
is the characteristic time scale associated to the ``resonance peak" of $\F_{\bf P}$; $Q\equiv \omega_{\rm max}\tau_{\max}$,
where $\omega\equiv \dot{\varphi}$ can be interpreted as the ``quality factor" associated to the same peak, 
and $\epsilon_{\rm max}\equiv \dot{\omega}_{\rm max}\tau^2_{\rm max}$. When $\bar{t}\gg \tau_{\rm max}$, 
the integrated recoil is analytically expected to be~\cite{Damour:2006tr}
\begin{align}
\label{eq:vend}
v^{\rm end}_{\rm A}\simeq \sqrt{2\pi} \F_{\bf P}^{\rm max}\dfrac{\tau_{\rm max}}{(1+\epsilon^2_{\rm max})^{1/4}}e^{-Q^2/[2(1+\epsilon^2_{\rm max})]}.
\end{align}
All relevant information to numerically evaluate Eqs.~\eqref{eq:vt}-\eqref{eq:vend} 
for NR (boldface) and RWZ$_\nu$ data is listed in Table~\ref{tab:CCC_max}. 
Several observations can be made. 
First, the presence of the antikick is {\it qualitatively} explained by the behavior 
of the complementary error function ${\rm erfc}(z)$, Eq.~\eqref{eq:vt},  
when $z$ is complex.  Since $\epsilon_{\rm max}$ is small, one sees that $\Im(z)$ 
is essentially given by $Q$~\cite{Damour:2006tr}.  
When $Q>0$ the usual, monotonic, behavior of $\rm{erfc}(z)$ 
is modified so that a local peak (the antikick) appears (see inset of Fig.~\ref{fig:antikick}). 
In particular, when $Q$ is small one finds  small or negligible antikicks; when $Q$ is larger 
the antikicks are larger.
Second, looking at the values of Table~\ref{tab:CCC_max} one sees that,
from the quantitative point of view the analytical result leads to estimates
of $v^{\rm end}_{\rm A}$ that are always systematically larger than the exact one, 
from $\sim 25\%$ ($q=2$) to $\sim 38\%$ ($q=\infty$). 
Third, focusing on the RWZ$_\nu$ data, from  Table~\ref{tab:CCC_max} 
one sees that the values of $\tau_{\rm max}$ and $Q$ do not vary much 
with the extrapolation with respect to the test-mass ones, 
contrary to $\F_{\bf P}^{\rm max}$, which is then the main responsible of getting 
$\hat{v}^{\rm end}_{\rm A}$ smaller than in the $\nu\to 0$ case. 
This gives a qualitative, analytical, consistency check of Table~\ref{tab:CCC} 
and Fig.~\ref{fig:fig_kick}. In addition, from Table~\ref{tab:CCC_max} 
one sees that $Q$ is always larger in the NR case than in the RWZ$_\nu$ one,  
which explains qualitatively Table~\ref{tab:CCC_antikick}. The reason for this 
is that the extrapolation acts only on the waveform modulus, and not on its phase 
(and frequency). As $Q=\omega_{\rm max}\tau_{\rm max}$, in the RWZ$_\nu$ case $\omega_{\rm max}$ 
is still driven by the underlying, less bound, dynamics of a particle on Schwarzschild spacetime, 
which, during late plunge and merger, spans frequencies  that are smaller 
than the corresponding (more bound) NR ones.
Similarly one explains the dependence of  $\Delta\hat{v}$ on $q$.

\section{Conclusions}
In the context of coalescing, nonspinning, black-hole binaries, 
we have found a simple way to correct the leading-order $\nu$-extrapolation of the 
recoil velocity in the test-mass limit, Eq.~\eqref{eq:Fitchett} (obtained via a 
perturbative approach) that is fully compatible with state-of-the-art numerical relativity 
simulations. Our approach is based on extrapolating in $\nu$ the test-mass waveform multipole 
by multipole using the corresponding leading-in-$\nu$ behavior before computing the recoil. 
An analogous $\nu$-extrapolation to get the final recoil velocity can be applied to the the waveform 
generated by a (spinning) particle plunging on a Kerr black hole. In this case, the subtlety 
is to {\it separately} extrapolate in $\nu$ the spin-dependent and the spin-independent 
part of the waveform because of their different, leading-order, $\nu$-dependence. 
The accuracy of the procedure will be discussed in future work.

\section*{Acknowledgements} 
I am indebted to S.~Bernuzzi for a discussion that inspired 
this work, and  to T.~Damour for constructive criticism. I thank A.~Zengino$\mathrm{\breve{g}}$lu  
for collaboration, and L.~Buchman, H.~Pfeiffer, M.~Scheel, B.~Szilagyi, J.~Gonzalez, 
B.~Br\"ugmann, M.~Hannam, S.~Husa, and U.~Sperhake for making available the data 
of their simulations. I acknowledge the Department of Physics, University of Torino, 
for hospitality during the development of this work. 

\bibliography{refs20131216.bib}{}

\end{document}